# Enhanced Mobile Computing Experience with Cloud Offloading

Hao Qian
hqianm@gmail.com

*Abstract*

The need for increased performance of mobile device directly conflicts with the desire for longer battery life. Offloading computation to multiple devices is an effective method to reduce energy consumption and enhance performance for mobile applications. Android provides mechanisms for creating mobile applications but lacks a native scheduling system for determining where code should be executed. This paper presents Jade, a system that adds sophisticated energy-aware computation offloading capabilities to Android apps. Jade monitors device and application status and automatically decides where code should be executed. Jade dynamically adjusts offloading strategy by adapting to workload variation, communication costs, and energy status in a distributed network of Android and non-Android devices. Jade minimizes the burden on developers to build applications with computation offloading ability by providing easy-to-use Jade API. Evaluation shows that Jade can effectively reduce up to 39% of average power consumption for mobile application while improving application performance.

Keywords — code offload; energy management; distributed computing; scheduling; mobile computing

## 1. Introduction

Mobile devices, such as smart phones and tablets, have become a necessity because they allow people to perform a wide variety of useful activities (e.g., video calls, emails, gaming, social networking, navigation, etc.) with mobility. These devices typically are equipped with a relatively powerful mobile processor, a rich set of sensors, and a substantial amount of memory. It allows previously unimaginable applications to be developed by integrating sensors such as motion sensors, position sensors, and environmental sensors.

However, battery life has become one of the biggest obstacles for mobile device advancements. Performance demanded by smartphones and tablets is increasing at a much faster rate than technological improvements in battery capacity. The need for increased performance of mobile devices directly conflicts with the desire for longer battery life.

One popular technique to reduce energy consumption of mobile devices is computation offloading in which an application reduces energy consumption by delegating code execution to other devices.[3,4,5] Traditionally, computations are offloaded to remote, resource-rich servers.[6] Selection of a proper offloading strategy can reduce power consumption and simultaneously enhance mobile device performance.

In this paper, we present Jade, an energy-aware computation offloading system for mobile devices. Jade, built for mobile devices running Android operating system, minimizes energy consumption of mobile applications through fine-grained computation offloading by providing the following benefits:

1. A runtime engine that enables computation offloading from mobile device to servers. By monitoring application and device status, the runtime engine automatically decides if code should run locally or remotely.
2. A programming model that helps developers create applications with computation offloading ability.

Our contributions are summarized as follows:
- We present the design and implementation of a complete system. Jade is able to offload computation from an Android device to any Android/non-Android devices.
- We present a multi-level scheduling algorithm that automatically and seamlessly transports workloads to the appropriate server based on performance and energy needs.
- We evaluate Jade with two applications. Results indicated that Jade can effectively reduce up to 39% of average power consumption for mobile device, while reducing the execution time of application.

## 2. System Design

In this section, we present the high-level design of Jade and its programming model in order to demonstrate how they integrate into one system, thereby supporting distributed execution of mobile applications.

For mobile applications with heavy computation needs, computation offloading is an effective method to reduce energy consumption and enhance performance.[8] However, it requires additional efforts and skills to develop applications with computation offloading ability and, unfortunately, no mature frameworks or tools exist for mobile application developers. We designed Jade to minimize the workload of developing applications with computation offloading ability by:

- Offering the Jade runtime engine which provides services for wireless communication, device profiling, program profiling and computation offloading. Conceptually, the Jade runtime engine automatically transforms application execution on one mobile device into a distributed execution optimized for wireless connection, power usage, and server capabilities.
- Providing an easy-to-use programming model for developers to build mobile applications that support energy-aware computation offloading.

In order to increase understanding of this offloading system, terms used in Jade must be defined. The mobile device that offloads computation is called *the client*. The device that receives and executes the offloaded code is called *the server*. Mobile applications contain *remotable tasks* which can be offloaded to the server (Figure 1). If a remotable task is executed on the client (i.e., it is not offloaded), we call it *local execution*. In contrast, if a remotable task is executed on the server (i.e., it is offloaded), we call it *remote execution*.

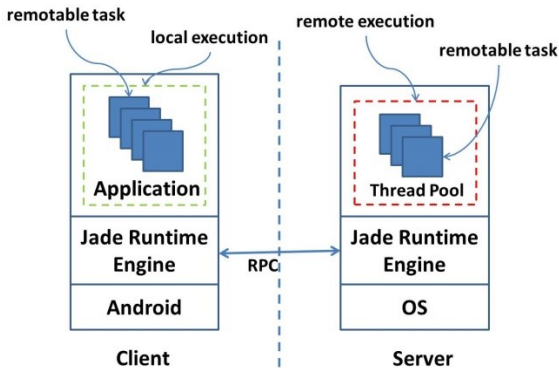

Figure 1: Jade enables computation offloading for mobile applications. Applications contain remotable tasks that can be offloaded from the client to the server. Jade runtime engine automatically decides where to execute remotable tasks and initiates distributed execution.

*2.1 Jade Runtime Engine*

In order to support computation offloading for applications, a computation offloading system is needed to handle some essential tasks such as:

- *Communication*. In order to offload code from the client to the server, the system should be able to 1) connect to other devices; 2) send data between devices; 3) coordinate with other devices for distributed execution; 4) track status of remote execution; 5) restore execution if unexpected errors occur (e.g., wireless connection lost, remote execution failure); and 6) exchange and record information related to all connected devices (e.g., connection speed, CPU usage, battery level, hardware configuration).
- *Profiling*. In order to make correct offloading decisions, the system should have fresh information regarding status of the device and application. Device profiling is the process of collecting information pertaining to device status, such as wireless connection, CPU usage, and battery level. Similarly, program profiling collects information about applications, such as execution time, energy consumption, memory usage, and data size.
- *Optimization*. The system should be able to determine an optimized offloading strategy in order to maximize application's energy savings and performance.

The goal of Jade is to maximize the benefits of energy-aware computation offloading for mobile applications, at the same time, minimizing the burden on developers to build such an application. By proving the Jade runtime engine, computation offloading is handled automatically in the background, allowing application developers to focus on the application without implementing the computation offloading mechanism.

Jade runtime engine components are shown in Figure 2. On an Android device, the Jade runtime engine runs as a group of background services. Jade supports two types of server: 1) *Android server* is device running Android and 2) *Non-Android server* is device running operating systems such as Windows and Linux. However, non-Android servers must have Java VM in order to support Jade, because Jade runtime engine runs as a Java program on a non-Android server.

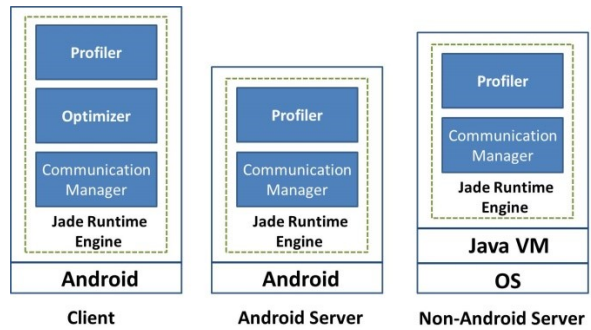

Figure 2: High-level design of the Jade runtime engine. Jade runtime engine supports Android/non-Android device.

Section 4 details the components of the Jade runtime engine.

*2.2 Jade Programming Model*

In addition to the Jade runtime engine, another key contribution of our system is the Jade programming model which helps developers efficiently build applications with computation offloading ability. Jade programming model provides APIs in order for mobile applications to interact with the Jade runtime engine. Use of the programming model allows developers to efficiently build applications that harness the power of computation offloading.

Details of the Jade programming model are provided in Section 4.

# 3. Multi-level Task Scheduling

Jade decides where to execute remotable tasks dynamically generated in an application (i.e., local execution or remote execution). Based on dynamic run time behavior of remotable tasks, such as execution time and energy consumption, remotable tasks should be offloaded to appropriate servers.

We implement a multi-level task scheduling algorithm in Jade that enables energy and performance-aware task scheduling. The algorithm 1) incorporates server status (e.g., type of power supply, computing capability, connection speed); 2) balances the workload between servers; and 3) offloads tasks to the most appropriate server according to the energy and computing demand of the task.

The multi-level task scheduling algorithm runs on the client and the server. Details are discussed in Sections 3.1 and 3.2.

*3.1 Task Scheduling on the Client*

The goals of task scheduling on the client include determining the appropriate server for each remotable task and balancing the workload for servers.

Tasks have varying computation demands. Tasks with heavy computation consume more energy, thereby requiring additional execution time. Servers also have varying characteristics. For example, mobile devices have less powerful hardware, such as CPU and RAM, and limited power supplies. Compared to mobile devices, desktop and laptop computers are often equipped with more powerful hardware and unlimited power supplies, making them better choices for tasks with heavy computation.

Server workload must also be considered. If a server currently operates with a high workload, it should not be a site to which new tasks are offloaded. In contrast, if a server is idle or has a light workload, it is able to receive new tasks.

We used Energy Delay Product (EDP) to measure computation demand of a remotable task. EDP, a performance measure that considers power and execution time, is defined as $EDP = T \times E = T^2 \times P$. For a remotable task $i$, T is execution time of $i$, E is energy cost to execute $i$, and P is average power consumption to execute $i$. Tasks with high computation demand should have longer execution time and consume more energy than tasks with low computation demand. Therefore, EDP can be used to classify tasks based on computation demand.

A distributed computing environment contains many useful information concerning execution parameters and performance that is readily available at each server but not readily available to the client. As a result, an information gap exists between the client and servers. In order to balance server workloads, the client ideally would have extensive information exchange with servers. Server status can change quickly, so the client must frequently gather information, thereby incurring many data transfers that consume energy. To reduce unnecessary energy costs, we used work stealing as a task scheduling strategy. Work stealing is a special case of task migration in which a "starving" device attempts to steal tasks from a "loaded" device. In Jade, the servers attempt to steal tasks from the client. The server monitors its own status and requests new tasks based on level of its current workload. For example, if the server was already overloaded, it would not ask for new tasks until remaining tasks were completed. The client does not need to monitor server status when using work stealing, so the amount of data transferred between devices is reduced, consequently saving energy.

On the client, two distinct buffers are used to store remotable tasks before offloading. One buffer is used for tasks with heavy computation (buffer H) and the other buffer is used for tasks with light computation (buffer L). When a remotable task $i$ is generated, Jade calculates the EDP value of $i$, and based on its EDP value, $i$ is put into buffer H or L. We also run benchmarking applications on every server to classify server as High Performance Device (H device) or Performance Constraint Device (C device). H devices steal tasks from buffer H and only steal from buffer L when buffer H is empty. Similarly, C devices steal tasks from buffer L and only steal tasks from buffer H when buffer L is empty. If only one kind of device is present (e.g., all servers are C devices), they steal tasks from both buffers (Figure 3).

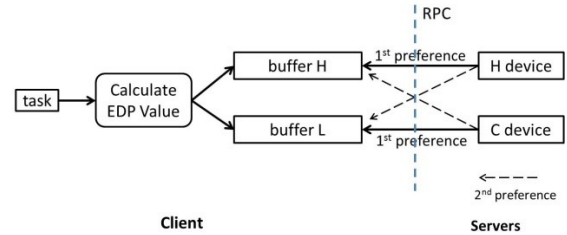

Figure 3: Servers stealing tasks from buffers of the client

*3.2 Task Scheduling on the Server*

Because mobile device users often spend limited time on one application, response time is very important for mobile applications. Given many similar applications to choose from, users do not use an application which feels sluggish, hangs, or freezes. To increase application responsiveness during computation offloading, the servers must return results of remote task execution as quickly as possible.

In the server, we used Highest Response Ratio Next (HRRN) algorithm to schedule tasks. In HRRN, the priority of each job depends on its estimated execution time and the amount of time the job has spent waiting. Jobs increase in priority as they wait, thereby preventing indefinite waiting (starvation). HRRN favors shortest jobs, but it also limits waiting time for long jobs. The priority of a task is defined as:

$$priority = \frac{waiting\ time + estimated\ run\ time}{estimated\ run\ time}$$

By selecting shorter tasks to execute, the client receives results frequently, making the application more responsive to the user.

# 4. Implementation

This section highlights important implementation details of Jade. Sections 4.1, 4.2, and 4.3 describe components in the Jade runtime engine. Section 4.4 shows details of the Jade programming model.

*4.1 Profiler*

When a remotable task is generated at run time, the Jade optimizer determines whether it should be executed locally or remotely. This decision is based on information provided by the profiler. The profiler collects the following information: 1) device status, such as charging status, battery level, CPU load, and wireless connection status); and 2) characteristics of remotable tasks, such as execution time, size, and energy consumption. The profiler measures characteristics of a remotable task during task's first execution and continuously monitors the status of devices because mobile device status changes frequently (e.g., loss of Wi-Fi connection, reaching low battery level). The optimizer may make wrong decision on whether the code should be offloaded based on a stale measurement.

Offline and online methods are available to measure power consumption of mobile devices and applications. An offline method, often used under laboratory conditions, uses external measurement tools. This method generates accurate results but requires special skills and equipment. However, most developers and users cannot realistically perform offline power measurement. The online method overcomes limitations of the offline approach by utilizing the Battery Monitoring Unit (BMU), which provides information related to battery (e.g., supply voltage, current, and battery temperature). A majority of current mobile devices are equipped with BMU. However, online method may not be as accurate as offline method because the information update rate of BMU is much lower than external measurement tools. According to previous research of mobile device power modeling, online method can generate satisfying results. The Jade profiler uses the online approach to measure energy consumption for devices and applications. The evaluation showed that Jade can make correct offloading decision when using information generated by this approach.

4.1.1 Program Profiling

When a new remotable task is generated, a unique name is assigned to it. (We combine the package name of the application and class name of the remotable task to form a unique name.) During the remotable task's first execution, the profiler collects the task's energy consumption $E$, execution time $T$, and size $S$. Then the information is provided to the optimizer. Based on the cost model, the optimizer decides if the task is suitable for offloading. Finally, all information concerning the remotable task is recorded in a database. When the same or similar task is executed again in the future, Jade will find its information in the database in order to understand whether or not the task should be offloaded.

With our automatic program profiling, developers only need to identify potential remotable tasks in a program and mark them according to the Jade programming model. When an application is installed on a mobile device, developers only need to run the application, invoking as many remotable tasks as possible. Jade automatically analyzes marked tasks in the background and decides whether or not they are remotable. Automatic program profiling greatly reduces the amount of work required to perform program power analysis, thus speeding up application development.

4.1.2 Device Profiling

At runtime, the profiler also continuously monitors device status. Collected information includes battery (i.e., battery level, charging status), wireless connection (i.e., Wi-Fi turned on/off, throughput), and CPU load.

Due to the nature of mobile devices, wireless connection could change frequently because a user could change locations. Fresh information about wireless connection is crucial in order for the optimizer to make correct offloading decisions. Similar to MAUI,[1] we used a simple method to measure the wireless link: Each time Jade offloads code, the profiler measures the transfer duration in order to obtain more recent average throughput. This simple approach takes into account both the latency and bandwidth characteristics of the network. We also built a simple energy cost model for wireless transfer using this approach: We send synthetic data from the client to the server, varying the size of the data, and we measure energy consumption of each transfer. This model allows us to predict energy consumption of data transfer as a function of data size.

*4.2 Optimizer*

The purpose of the Jade optimizer is to choose suitable remotable tasks to offload to the server in order to find an offloading strategy that minimizes the application's energy consumption. The optimizer makes the offloading decision by solving an optimization problem using information provided by the profiler.

Characteristics of a remotable task determine where it should be executed. The code should be offloaded only if the energy cost to locally execute it is greater than the energy cost required to transfer it. For example, code performing heavy computation on small data is suitable for offloading. For some code, the energy cost of transfer outpaces the energy cost of local execution (e.g., lightweight computation on big data), such code should not be offloaded.

Based on information provided by the profiler, the optimizer determines the best offloading strategy by solving the following cost model. $I$ represents the set of remotable tasks in the application. For each remotable task $i \in I$, $E_i^e$ is energy consumed to execute $i$ locally. $E_i^t$ is energy consumed to transfer $i$ between the client and the server. $T_i^l$ is the execution time of $i$ on the client. $T_i^r$ is the execution time of $i$ on the server. $T_i^t$ is the transfer time of $i$. $I_i$ is the indicator variable: $I_i = 0$ if $i$ is executed locally, $I_i = 1$ if $i$ is executed remotely. The optimizer must find the assignment for $I_i$ such that:

- *maximizes* $\sum_{i \in I} I_i \times (E_i^e - E_i^t)$

- guarantees $\sum_{i \in I}(1 - I_i) \times T_i^l + I_i \times T_i^r + I_i \times T_i^t \leq l$

The first formula is total energy savings. The second constraint stipulates that the total execution time is within $l$. Developers can configure $l$ according to their specific requirements.

As explained in Section 4.4, the Jade programming model requires that every remotable task $i$ to perform independent job, so no dependencies exist between tasks. This requirement further simplifies the cost model. For each remotable task $i$, the optimizer must solve only the following inequation:

$$E_i^e - E_i^t > 0$$
$$T_i^r + T_i^t - T_i^l \leq l$$

$E_i^e - E_i^t$ is the energy saving if $i$ is executed remotely. $i$ should be considered for offloading only when $E_i^e > E_i^t$. Similarly, the second inequation guarantees that time difference between remote execution and local execution is not greater than $l$.

*4.3 Communication Manager*

Code offloading is handled by the communication manager. The communication manager is responsible for code manipulation (i.e., code transfer, code execution) and device coordination.

The communication manager handles code offloading by following these steps:
1. Records code information in the offloaded code table (Table 1). The purpose of the table is to track the status of offloaded code.
2. Offloads the code to the server.
3. The server receives the code and executes it in a new thread.
4. The server sends the code back to the client after execution is complete.
5. The client receives the code and updates code information in the offloaded code table.

Table 1: Example of offloaded code table

| ID | Offloaded | Server | Returned | Result |
|---|---|---|---|---|
| 0001 | true | 192.168.49.1 | true | finish |
| 0002 | true | 192.168.49.1 | false | |
| 0003 | true | 192.168.49.1 | true | fail |

For remote execution failure, the communication manager has mechanisms to guarantee correctness of the application. For example, if wireless connection is lost during remote execution, the client's communication manager re-executes the code locally and the server's communication manager abandons the execution. If the returned code shows its result as failed, it is also re-executed on the client.

*4.4 Jade Programming Model*

When designing applications that support code offloading, the application must be partitioned into sub-parts which can be offloaded to the server. Applications can be partitioned at various levels, such as class, method, process, and thread. An application is partitioned at the class level in Jade. Developers can produce an initial partition of their applications with minimal effort using the Jade programming model.

To be considered for offloading, a class must implement one of two interfaces: *RemotableTask* interface or *RemotableGenTask* interface. As mentioned, Jade supports Android and non-Android servers. If a class contains Android code, it must run on Android devices, so the class must implement *RemotableTask* interface which is guaranteed to be offloaded to an Android server. If a class contains only Java code which does not access any Android API, then the class should implement *RemotableGenTask* interface, Jade can offload such class to more servers (i.e., any server with Java VM). With the exception of different target servers, *RemotableGenTask* and *RemotableTask* have similar mechanisms that support code offloading. Therefore, we used *RemotableTask* interface to demonstrate how it works.

In Jade, a class that implements the *RemotableTask* interface is called a *remotable class*. An instance (object) of *remotable class* is called a *remotable object*. A *remotable object* can be executed on the client (locally) or on the server (remotely). An application developed using the Jade programming model is called a Jade compatible application. At runtime, a Jade compatible application contains *remotable objects* that can run concurrently on multiple servers.

Some types of code must be executed locally. If a class contains the following code, it should not be considered for offloading:
- Code that creates user interface of the application.
- Code that handles user interaction (e.g., callback method for clicking a button).
- Code that accesses the client's special hardware which may be unavailable on the server (e.g., some smart phones are equipped with GPS sensor, but most computers are not).
- Code unsuitable for re-execution (e.g., code that performs online transactions).[1]

Our goal in providing the *RemotableTask* interface is to eliminate the need for developers to guess whether or not a class is suitable for offloading in terms of energy consumption and performance. When a class does not contain the above code, it can implement the *RemotableTask* interface, and the Jade runtime engine automatically determines if the class is suitable for offloading.

The *RemotableTask* interface is the key construct in the Jade programming model, it defines the following methods called by the Jade runtime engine in sequence:
- *preExecution*, performs preparations before executing the main task (e.g., connect database, initialize data).

- *loadData*, loads data from the client's file system. If *loadData* is called on the server, data is read and transferred from the client.
- *execution*, performs the main task.
- *updateData*, the counterpart of *loadData*. After the task finishes, *updateData* updates data in the client's file system. If *updataData* is called on the server, data is transferred back to the client.
- *postExecution*, the counterpart of *preExecution* that performs remaining tasks after the main task is complete (e.g., disconnect database, update log and send notification to user).

By implementing the RemotableTask interface, the life of a remotable object is divided into three stages (Figure 4): before execution, execution, and after execution. This division naturally matches steps of a common computation: 1) preparation (e.g., load data, connect to network); 2) execution, which performs computation on the data; and 3) update, which finishes the execution (e.g., update database, notify user).

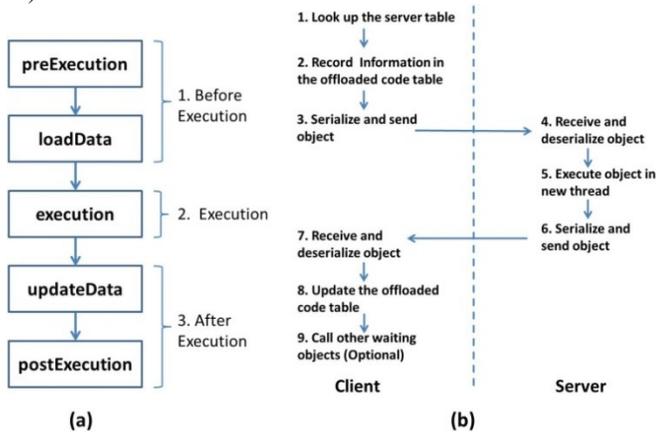

Figure 4: (a) Life of a *remotable object* is divided into three stages by implementing the *RemotableTask* interface (b) Code offloading workflow

The Jade optimizer considers all *remotable objects* of an application for offloading. If a *remotable object* is eligible for offloading, the Jade runtime engine handles the code offloading by following the workflow shown in Figure 4.

Application development using Jade programming model is intuitive and simple, developers only need to follow two steps:
1. Identify tasks in the application that can be offloaded (code does not violate previously mentioned rules).
2. For each task, create a class that implements the *RemotableTask* interface.

For example, an application can potentially perform three tasks: 1) collect information entered by the user (e.g., name, phone number, address and company); 2) verify the information; and 3) save the information to the database on a remote server. Task three is a good candidate for computation offloading. Using the Jade programming model, we create a remotable class (UpdateInfo) for Task three. The pseudocode is:

*class UpdateInfo implements RemotableTask{*

```
preExecution(){
   open database connection;
}
loadData(){
   do nothing;
}
execution(){
   if new user
      insert user info into database;
   else
      update user info;
}
updateData(){
   do nothing;
}
postExecution(){
   close database connection;
   send notification to user;
}
}
```

In Jade, a *remotable object* should perform tasks independently with no dependencies between *remotable objects*. Advantages of this requirement include 1) reduction of cost model complexity so the optimizer consumes less energy and 2) scalable remote task offloading in which the number of offloaded tasks can dynamically scale up/down with adding/removing of servers.

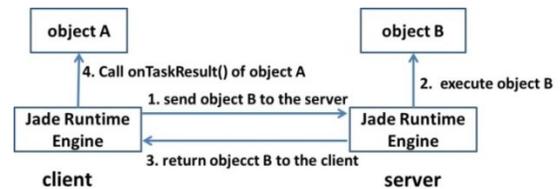

Figure 5: Jade programming model provides a mechanism for object B to notify object A when object B finishes execution. Object A must implement the *OnTaskReturnedListener* in order to receive the notification.

The Jade programming model also provides a mechanism for a *remotable object* to notify local objects when the *remotable object* finishes execution (Figure 5). If a class needs to be notified, the class can implement the *OnTaskReturnedListener* interface. This interface defines the *onTaskResult* callback method called by the Jade runtime engine when a specified *remotable object* finishes execution.

## 5. Evaluation

In this section we evaluate Jade's ability to reduce energy consumption for mobile applications. We implemented two applications that perform common tasks widely used by users.

Image processing, such as image editing and object recognition, is used in many applications. Our first application was FaceDetection (FD) which asks users to select pictures from the device's photo gallery. FD then detects human faces in those pictures and highlights them by putting a rectangle around each face. Code performing face detection was implemented in a remotable class in order to be offloaded to the server.

Navigation application is one of the most popular mobile applications. Path finding in such applications could be computation intensive, making it suitable for computation offloading. Our second application was FindRoute (FR). This application simulates a real world navigation application by determining the shortest path between two nodes of a graph. We used Dijkstra's algorithm for path finding, and path finding code was implemented in a remotable class.

We used a Moto X smart phone as the client. Moto X is equipped with Qualcomm Snapdragon 1.7GHz Dual-Core CPU, quad-core GPU, and 2GB RAM. We used a Samsung Galaxy S3 smartphone and a Dell Inspiron 15 laptop as the servers. Galaxy S3 has a Quad core 1.4GHz processor and 1GB RAM. Inspiron 15 has a 1.8GHz Dual-Core CPU and 8GB RAM. All devices were connected in the same wireless network.

In order to evaluate how much energy Jade can save for mobile device, we run FD and FR on Moto X. FD performs face detection on 50 pictures, the size of each picture is under 200KB. Each application was executed twice: first time with Jade disabled and second time with Jade enabled. We compared the first execution with the second execution to evaluate the performance of Jade.

Results showed that Jade effectively reduces power consumption for FD (Figure 6). Average power consumption was reduced by 34% (Figure 10). FR had similar results (Figure 8): Jade reduced 39% of average power consumption for FR. Power reduction was achieved because when Jade was enabled, some computations were offloaded to servers, thereby, reducing the workload of the client (Figure 7 and 9). Because more tasks were executed concurrently, Jade also improved the performance of both applications: it reduced 37% and 45% of execution time for FD and FR respectively (Figure 11).

Results demonstrated that Jade can effectively reduce energy consumption of mobile applications and improve application performance.

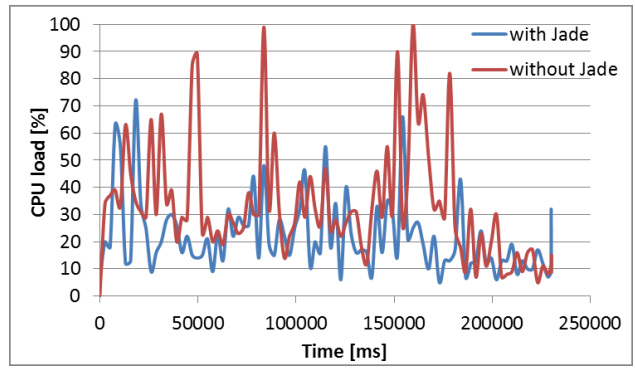

Figure 7: CPU load of FaceDetection

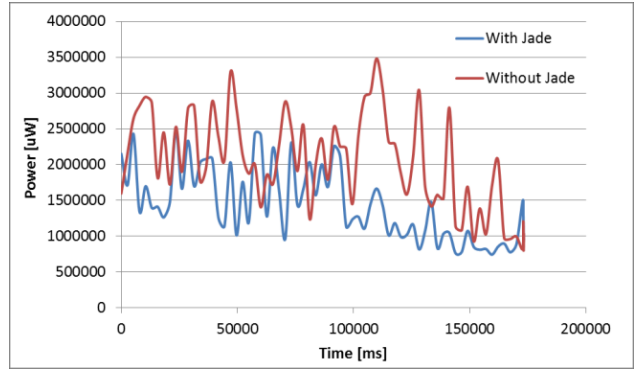

Figure 8: Power consumption of FindRoute

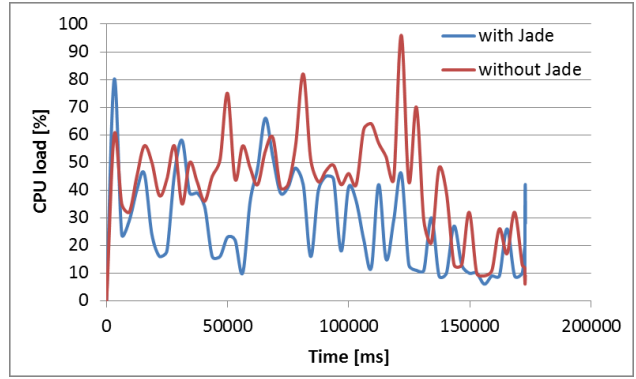

Figure 9: CPU load of FindRoute

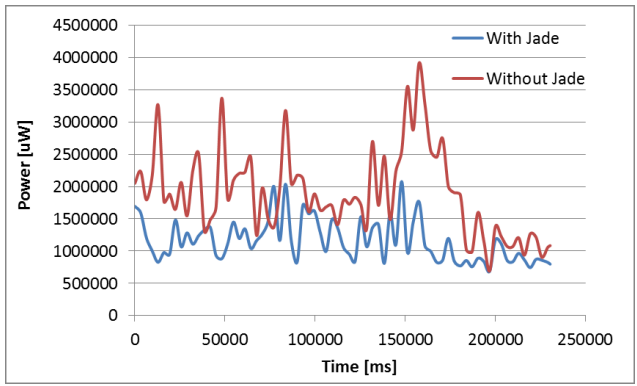

Figure 6: Power consumption of FaceDetection

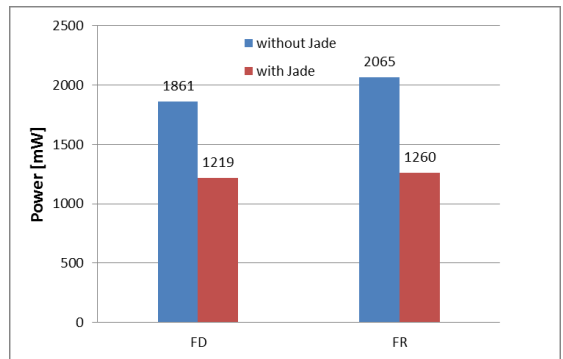

Figure 10: Average power consumption of FaceDetection and FindRoute

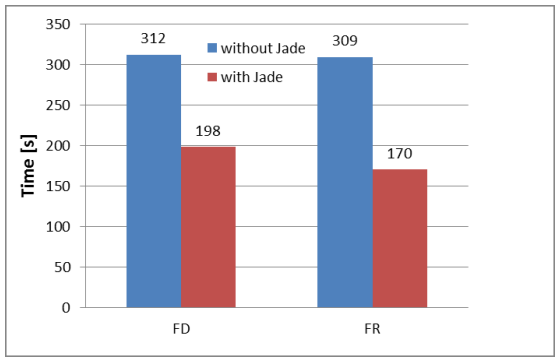

Figure 11: Execution time of FaceDetection and FindRoute

## 6. Related Work

Mobile devices have limited resources such as battery capacity, storage, and processor performance. Computation offloading is an effective method to alleviate these restrictions by sending heavy computations to resourceful servers and receiving results from these servers. Many issues related to computation offloading have been investigated in the past decade, including feasibility of offloading, offloading decisions, and development of offloading infrastructures.

Jade was built upon previous research regarding program partitioning, code offloading, and remote execution. In this section, we provide an overview of proposals by these researches and how they relate to Jade.

Cuervo et al. proposed MAUI,[1] a system that enables energy-aware offloading of mobile code to the infrastructure. MAUI enables developers to produce an initial partitioning of their applications by annotating methods and/or classes as remotable. At runtime, the MAUI solver decides which remotable methods should execute locally and which should execute remotely. Unlike MAUI, Jade provides a sophisticated programming model with a full set of APIs, so developers have total control on: how application is partitioned, where code is offloaded and how remotable code interacts with local code. In Jade, dependencies do not exist between remotable tasks, the profiler and optimizer do not need to analyze the whole program, thereby, energy cost of program profiling and cost model calculation is lower than MAUI.

Chun et al. proposed CloneCloud,[2] an application partitioner and execution runtime that enables unmodified mobile applications running in an application-level virtual machine to seamlessly offload part of their execution from mobile devices onto device clones operating in a computational cloud. In CloneCloud, threads must be paused, all states of the threads must be transferred to the server, and then threads resume on the server in order to offload computation. Offloading is expensive, however, especially when the client and server are both resource constraint mobile devices. In contrast, code offloading in Jade is lightweight. Remotable objects are serialized, transferred, and deserialized, resulting in much lower overhead compared to thread migration.

## 7. Future Work

In our future work, we will extend Jade to include cloud platform. Cloud provides a scalable and powerful computing environment which is ideal for complex computing tasks. Today, many mobile devices have fast wireless link such as 4G LTE without limitation of locations, making cloud platform a good destination for computation offloading. We will also study how cloud changes the development pattern of mobile applications.

Another direction of future work is security. Jade offloads computation to trusted devices, but security concerns arise if computation offloading is scaled up, for example, multiple applications accessing a single server, running foreign code on the server, and remote codes interfering with each other.

## 8. Conclusion

In this paper, we presented Jade, a system that enables computation offloading for mobile devices. Jade can effectively reduce energy consumption of mobile devices, and dynamically change its offloading strategy according to device status.

We evaluated Jade with two applications: a face detection application and a path finding application. Results showed that Jade can effectively reduce energy consumption for both applications while improving their performance.